# Modified Quine-McCluskey Method

Jadhav Vitthal, Buchade Amar,
Pune, Maharashtra, India.
{jadhavvitthal1989, amar.buchade}@gmail.com

## ABSTRACT

The digital gates are basic electronic component of any digital circuit. Digital circuit should be simplified in order to reduce its cost by reducing number of digital gates required to implement it. To achieve this, we use Boolean expression that helps in obtaining minimum number of terms and does not contain any redundant pair. Karnaugh map (K-map) and Quine-McCluskey (QM) methods are well known methods to simplify Boolean expression. K-map method becomes complex beyond five variable Boolean expression. Quine-McCluskey method is computer based technique for minimization of Boolean function and it is faster than K-map method. This paper proposes E-sum based optimization to Quine-McCluskey Method to increase its performance by reducing number of comparisons between mintermlist in determination of prime implicants. Modified Quine-McCluskey method (MQM) can be implemented to any number of variable.

## 1. INTRODUCTION

The English mathematician and philosopher George Boole invented the Boolean algebra in 1854. A few decades later C.E. Shannon showed how the Boolean algebra can be used in the design of digital circuits (Shannon, 1938).Digital gates are basic component of any digital circuit. Boolean logic is the basic concept that underlies all modern electronic digital computers. Using Boolean laws it is possible to minimize digital logic circuits by reducing the original number of digital components (gates) required to implement digital circuits. This will reduce the chip size, the cost and increases the speed of circuit. Since minimization with the use of Boolean laws is neither systematic nor suitable for computer implementation, a number of algorithms were proposed in order to overcome the implementation issue. Karnaugh proposed a technique for simplifying Boolean expressions using an elegant visual technique, which is actually a modified truth table intended to allow minimal sum-of products (SOP) and product-of-sums (POS) expressions to be obtained (Karnaugh, 1953). The Karnaugh Map (K-Map) based technique breaks down beyond six variables. Quine and McCluskey proposed an algorithmic based technique for simplifying Boolean logic functions (McCluskey-1956, Quine-1952).

The Quine-McCluskey (QM) method is a computer-based technique for Boolean function simplification and has mainly two advantages over the K-Map method. Firstly, it is systematic for producing a minimal function that is less dependent on visual patterns. Secondly, it is a viable scheme for handling a large number of variables. But it is difficult to work manually with Quine-McCluskey method when number of variable increases in the minterm. This paper proposes the alternative principal for grouping which simplifies the process of grouping in QM using E-sum (Elimination sum).

## 2. DEFINITION

### 2.1 Sum of Product (SOP)
Sum of product (SOP) is more common form of Boolean expression. The expression are implemented as AND gates (product) feeding a single OR gate (sum).

### 2.2 Product of Sum (POS)
Product of sum (POS) is less commonly used form of Boolean expression. The expression are implemented as OR gates (sum) feeding a single AND gate (product).
E.g. $Y = (A + B)(A + B + C)(\overline{A} + \overline{B})$ is in POS form.

### 2.3 Canonical Form
The Boolean expression are said to be in canonical or standard form if each term contain all available input variable.
E.g. Suppose $Y = AB + ABC + \overline{A}\overline{B}$, Then its equivalent canonical form is given as below.

$$Y = AB(C+\overline{C}) + ABC + \overline{A}\overline{B}(C+\overline{C}) \quad (\because C + \overline{C} = 1)$$
$$= ABC + AB\overline{C} + \overline{A}\overline{B}C + \overline{A}\overline{B}\overline{C}$$

### 2.4 Mintermlist
It is the set of minterm which can be combined.
E.g. (1, 5) and (0, 1, 2, 3) are mintermlist.

### 2.5 Prime Implicant
It is the mintermlist which cannot be combined with any other mintermlist.

### 2.6 E-sum (Elimination sum)
Sum of eliminated variable's positional weight in mintermlist is called as E-sum. Thus E-sum is used to keep track of eliminated variable's position in mintermlist.
E.g. Let $\overline{AB}--$ be mintermlist in which eliminated variable is denoted by '-'. In this example variable C and D are eliminated and 1,2 are the positional weights of variable C and D respectively. Hence E-sum= 1+ 2 =3.

## 3. MATCHING PRINCIPAL OF MQM

Two mintermlist in adjacent group are said to be matched or combined if their 'E-sum' is equal and ' (least minterm in a mintermlist of $(n + 1)^{th}$ group) - (least minterm in a mintermlist of $n^{th}$ group) ' produce valid positional weight for binary number system ( i.e.1,2,4,8,16…). This difference acts as 'current mismatch positional weight' (MPW) for resulting mintermlist. Thus to combine two mintermlist they must satisfy below two condition.

1. Least minterm in a mintermlist of $(n+1)^{th}$ group) - (least minterm in a mintermlist of $n^{th}$ group) = $2^n$ where n >= 0 .

2. Elimination sum of both mintermlist must be equal.

## 4. MQM ALGORITHM

The proposed algorithm uses E-sum and algebraic approach to reduce number of comparisons between mintermlist. E-sum is used to keep track of all eliminated variable in mintermlist or Boolean term. E-sum based MQM algorithm is presented using following step-by-step approach:

1. Transform given Boolean function into canonical SOP form and obtain binary notation for each minterm.

2. Arrange all minterm into groups according to number of 1's in their binary notation. All minterms in one group should contain same number of 1's. Then initialize E-sum of all minterms to 0.

3. Compare mintermlist in adjacent groups according to MQM matching principal. Use algebraic approach to reduce number of comparison between mintermlist in adjacent group. In Algebraic approach mintermlist in $n^{th}$ group having least minterm as x is compared with all mintermlist in $(n+1)^{th}$ group having least minterm as $x + 2^P$ Where P = 0,1,2,3…. so on. Once there are any two mintermlist of $n^{th}$ and $(n+1)^{th}$ group satisfying MQM matching principal and having least minterm as x and y respectively. Then combine the two mintermlist by taking E-sum of resulting mintermlist as "E-sum of combining mintermlist + Current MPW (i.e. y – x )". Checkmark (√) mintermlists which can combined to form new mintermlist. Now repeat the same procedure for all other mintermlists.

4. Eliminate repeated or identical mintermlist from combined mintermlist in all group. Two mintermlist in same group become identical if their corresponding E-sum, least minterm and largest minterm are equal.

5. Repeat step described in 4.3 and 4.4 to minimize given Boolean function until it is impossible to combine mintermlist.

6. Collect all non-checked (i.e. not √ marked) mintermlists as prime implicant.

7. Now eliminate the redundant prime implicants by using Prime implicant chart as in Quine-McCluskey method.

## 5. EXAMPLE

**Example 1** Minimize the following Boolean function

$F(A, B, C, D) = \sum m(4,5,6,8,9,10,13) + d(0,7,15)$

1. Table 1

| Mintermlist | Binary Notation A B C D |
|---|---|
| 0 | 0 0 0 0 |
| 4 | 0 1 0 0 |
| 5 | 0 1 0 1 |
| 6 | 0 1 1 0 |
| 7 | 0 1 1 1 |
| 8 | 1 0 0 0 |
| 9 | 1 0 0 1 |
| 10 | 1 0 1 0 |
| 13 | 1 1 0 1 |
| 15 | 1 1 1 1 |

2. Arranging all minterm into groups according to number of 1's in their binary notation and initializing E-sum of all mintermlist to zero.

Table 2

| Group No. | Mintermlist | Elimination Sum (E-sum) | |
|---|---|---|---|
| 0 (No one's) | 0 | 0 | √ |
| 1 (One one's) | 4 | 0 | √ |
| | 8 | 0 | √ |
| 2 (Two one's) | 5 | 0 | √ |
| | 6 | 0 | √ |
| | 9 | 0 | √ |
| | 10 | 0 | √ |
| 3 (Three one's) | 7 | 0 | √ |
| | 13 | 0 | √ |
| 4 (Four one's) | 15 | 0 | √ |

3. Now we have to compare E-sum of $n^{th}$ group's mintermlist having least minterm as x with E-sum of $(n+1)^{th}$ group's mintermlists starting with $x + 2^P$ where P = 0,1,2,3,…. So on.

   **1. Comparison between Group 0 and Group 1:**
   Mintermlist 0 is compared with mintermlist 4 $(0 + 2^2)$, 8 $(0 + 2^3)$. E-sum of both mintermlist 0 and 4 are equal. So They can be combined to form mintermlist- (0, 4) with E-sum (0, 4) = E-sum (0) + (4-0) = 4. Similarly mintermlist-0 and mintermlist-8 can be combined as (0, 8) with E-sum = 8.

   **2. Comparison between Group 1 and Group 2 :**
   Mintermlist 4 is compared with mintermlist 5$(4 + 2^0)$, mintermlist 6 $(4+2^1)$ which result into mintermlist (4,5) with E-sum=1 and mintermlist (4,6) with E-sum=2. Similarly by comparing mintermlist 8 with mintermlist 9 $(8+ 2^0)$ and 10 $(8 + 2^1)$, we get mintermlist (8, 9) with E-sum=1 and mintermlist (8, 10) with E-sum=2.

**3. Comparison Between Remaining Adjacent Group :**
   Similarly when we compare mintermlist in remaining adjacent group, we get combined mintermlist as (5,7), (5,13), (6,7), (9,13), (7,15), (13,15) with E-sum as shown in Table 3. Note that all mintermlist which can be combined are check-marked (i.e. √ marked) in Table 2.

**Table 3**

| Group No. | Mintermlist | Elimination Sum (E-sum) |
|---|---|---|
| 0 | (0, 4)<br>(0, 8) | 4 (0 + 4)<br>8 (0 + 8) |
| 1 | (4,5)<br>(4,6)<br>(8,9)<br>(8,10) | 1 (0 + 1) √<br>2 (0 + 2) √<br>1 (0 + 1)<br>2 (0 + 2) |
| 2 | (5,7)<br>(5,13)<br>(6,7)<br>(9,13) | 2 (0 + 2) √<br>8 (0 + 8) √<br>1 (0 + 1) √<br>4 (0 + 4) |
| 3 | (7,15)<br>(13,15) | 8 (0 + 8) √<br>2 (0 + 2) √ |

**4** There is no repeating identical mintermlist in Table 3.

**5**
1. Repeating process in step 4.3 of algorithm, we get 'quad of minterm' as shown in Table 4. Mintermlist of Table 3 which can be combined are check-marked in Table 3.

**Table 4**

| Group No. | Mintermlist | E-sum |
|---|---|---|
| 0 | | |
| 1 | (4,5,6,7)<br>(4,6,5,7) | 3 (1 + 2)<br>3 (2 + 1) |
| 2 | (5,7,13,15)<br>(5,13,7,15) | 10 (2 + 8)<br>10 (8 + 2) |

2. Now in above Table 4 mintermlist (4,5,6,7) and (4,6,5,7) are identical as they have same least minterm (i.e. 4), largest minterm (i.e. 7) and E-sum (i.e. 3). Similarly mintermlist (5,7, 13, 15) and (5, 13,7, 15) are identical. Now modify Table 4 by excluding one of the identical mintermlist from above Table 4 to eliminate redundancy.

**Modified Table 4**

| Group No. | Mintermlist | E-sum |
|---|---|---|
| 0 | | |
| 1 | (4,5,6,7) | 3 (1 + 2) |
| 2 | (5,7,13,15) | 10 (2 + 8) |

3. In above Table Mintermlist (4,5,6,7) and (5,7,13,15) cannot be combined due to their unequal E-sum. So process of grouping end here.

**6** By collecting all unchecked (i.e. not √ marked) mintermlist from Table 1, Table 2, Table 3 and modified Table 4, we get prime implicant as (0,4), (0,8), (8,9), (8,10), (9,13), (4,5,6,7), (5,7,13,15).

**7** Now by using Prime Implicant chart we can eliminate redundant mintermlists (0,4),(0,8),(8,9), (5,7,13,15).

Thus $F(A,B,C,D) = \underline{(8,10)} + \underline{(9,13)} + (4,5,6,7)$
$= A\overline{B} + A\overline{B}D + \overline{A}C\overline{D}$

## 6. COMPARISON BETWEEN QM AND MQM IN FIRST PASS FOR WORST CASE

In Quine-McCluskey or Modified Quine-McCluskey method number of comparison between two mintermlist in adjacent group increases with increase in number of minterm in group. In case of QM or MQM worst case is the case in which maximum no of comparison are required or Boolean function contain all possible combination of input. So in worst case Boolean function minimizes to logic 1. N variable Boolean function contains $2^n$ minterms in worst case. QM or MQM method work in passes. Maximum N passes are required to minimize N variable Boolean function. In $i^{th}$ pass mintermlists in adjacent group containing $2^{i-1}$ minterms are compared.

Each minterm in $i^{th}$ group is compared with each minterm in $(i+1)^{th}$ group in QM method. QM or MQM method has $^nC_i$ minterm in $i^{th}$ group for worst case. So total number of comparison between mintermlist of adjacent group in first pass is given by below formula.

Total No. of comparisons $= \sum_{i=0}^{n-1} {^nC_i} * {^nC_{i+1}}$ (1)

For four variable Boolean function substituting n = 4 in (1), we get Total No. of comparisons $= {^4C_0} * {^4C_1} + {^4C_1} * {^4C_2} + {^4C_2} * {^4C_3} + {^4C_3} * {^4C_4} = 56$. But in case of MQM each minterm in $i^{th}$ group is compared with (n-i) minterm in $(i+1)^{th}$ group. So Total No. of comparison between mintermlist in adjacent group for n variable Boolean function in first pass is given by below formula.

Total No. of comparisons $= \sum_{i=0}^{n-1} {^nC_i} *(n-i) = n*2^{n-1}$ (2)

For four variable Boolean function number of required comparison between mintermlist are 32 (from Table 5). This can be verified with help of (2). By substituting n=4 in (2), we get total number of comparisons between mintermlist in adjacent group $= {^4C_0}*4 + {^4C_1}*3 + {^4C_2}*2 + {^4C_3}*1 = 4+12+12+4=32$.

**Table 5**

| Group No. | Mintermlist | Compared with | No. of comparison |
|---|---|---|---|
| 0 (No one's) | 0 | 1,2,4,8 | 4 |
| 1 (One one's) | 1 | 3,5,9 | 3 |
|  | 2 | 3,6,10 | 3 |
|  | 4 | 5,6,12 | 3 |
|  | 8 | 9,10,12 | 3 |
| 2 (Two one's) | 3 | 7,11 | 2 |
|  | 5 | 7,13 | 2 |
|  | 6 | 7,14 | 2 |
|  | 9 | 11,13 | 2 |
|  | 10 | 11,14 | 2 |
|  | 12 | 13,14 | 2 |
| 3 (Three one's) | 7 | 15 | 1 |
|  | 11 | 15 | 1 |
|  | 13 | 15 | 1 |
|  | 14 | 15 | 1 |
| 4 (Four one's) | 15 |  | 0 |
| **Total Number of Comparison between mintermlists in adjacent group** | | | **32** |

Table 6 shows number of comparison between mintermlist required to minimize n variable Boolean function for QM and MQM method in worst case. QM method require 11440 comparisons which is 11.17 times greater than comparisons needed in MQM method (1024) to minimize 8 variable Boolean function in worst case. Thus MQM method is faster than QM method.

**Table 6**

| No. of Variable (n) | Total Comparisons in worst case | |
|---|---|---|
|  | QM Method $\left(\sum_{i=0}^{n-1} {}^nc_i * {}^nc_{i+1}\right)$ | MQM Method $(n*2^{n-1})$ |
| 2 | 4 | 4 |
| 3 | 15 | 12 |
| 4 | 56 | 32 |
| 5 | 210 | 80 |
| 6 | 792 | 192 |
| 7 | 3003 | 448 |
| 8 | 11440 | 1024 |

## 7. DISADVANTAGES OF MQM

The Boolean expression or minterm which is used as input to MQM or QM must be in standard SOP form. So we have to convert Boolean expression into standard SOP form before using MQM. This conversion process adds redundant minterm in input.

E.g. Let Y = AD + ABC . Now Boolean expression must be converted to standard SOP form to use MQM method.

i.e. Y = AD + ABC
= A(B+$\bar{B}$) (C+$\bar{C}$) D + ABC (D+$\bar{D}$)   (∵ A+$\bar{A}$ = 1)
= ABCD + AB$\bar{C}$D + A$\bar{B}$CD + A$\bar{B}$ $\bar{C}$D + ABC$\bar{D}$

Here number of minterm in input is increased by 3 which will increase number of comparison required in grouping process. This will increase time complexity of MQM.

## 8. CONCLUSION

In this paper a new method called Modified Quine-McCluskey (MQM) method is introduced by using which performance of digital circuit can be increased by reducing number of literal in Boolean function. Algebraic approach is used to reduce number of comparison between mintermlist while E-sum is used to eliminate repetition. MQM is simple and faster than Quine-McCluskey method due to less number of required comparisons. Thus method can be used to achieve speed in minimizing the Boolean function manually and to improve performance of conventional method.